\documentclass[twocolumn, prx, aps, superscriptaddress, longbibliography, showpacs, amsmath, amssymb, floatfix]{revtex4-1}
\usepackage[colorlinks,linkcolor=blue,anchorcolor=blue,citecolor=blue,urlcolor=blue]{hyperref}
\usepackage{graphicx}
\usepackage{amssymb}
\usepackage{amsmath}
\usepackage{epsfig}
\usepackage{color}
\usepackage{mathtools}
\usepackage{bm}
\usepackage{bbm}
\usepackage{physics}
\renewcommand{\vec}[1]{\bm{#1}}
\begin{document}
\title{A new theory bridging non-relativistic and QED-based path integrals unveils more than quantum mechanics}
\author{Wei Wen}
\email{wenwei@hut.edu.cn}
\affiliation{College of Science, Hunan Univ ersity of Technology, Zhuzhou 412007, Hunan Province, China}

\date{\today }

\begin{abstract}
The Feynman path integral plays a crucial role in quantum mechanics, offering significant insights into the interaction between classical action and propagators, and linking quantum electrodynamics (QED) with Feynman diagrams. However, the formulations of path integrals in classical quantum mechanics and QED are neither unified nor interconnected, suggesting the potential existence of an important bridging theory that could be key to solving existing puzzles in quantum mechanics. In this work, we delve into the theoretical consistency, completeness, and integration with established path integral theories, revealing this concealed path integral form. This newly uncovered form not only connects various path integral approaches but also demonstrates its potential in explaining quantum phenomena like the origin of spin and quantum nonlocal correlations. It transcends conventional quantum mechanics, proposing a more profound and fundamental physical principle.
\end{abstract}

\maketitle
\section{Introduction}
The Feynman path integral is a pivotal representation in quantum mechanics and a vital method in quantum field theory studies. It establishes a relationship between quantum mechanics and classical mechanics, explaining the impact of particle trajectories on propagators and indirectly supporting the principle of least action in classical mechanics. Since its inception in 1948\cite{Feynman1948}, the path integral formulation has catalyzed some scholarly discourse and investigative pursuits. It ingeniously bridges the non-relativistic classical action, $S_c$, with the Schr\"odinger equation in non-relativistic quantum mechanics. However, it conspicuously lacks a parallel formulation for incorporating the relationship between relativistic classical action, $S_r$, with Dirac equation in relativity quantum mechanics. Moreover, the QED-based path integral theory does not have any correspondence to classical actions, implying its departure from classical analytical mechanics\cite{Wharton2015}. Despite extensive efforts by researchers to reconcile quantum field theory with a classical interpretative framework, a significant theoretical divide persists\cite{BARUT19891,PhysRevD.45.2044,BARVINSKY1998533,Seide2006,Johnson2010,PhysRevLett.89.250403,KULL2002147,Kull:1999aa,PhysRevLett.53.419}.

Currently, the Feynman path integral theory and the path integral formulation under QED remain disjointed, with the former not being considered a low-energy approximation of the latter and the latter not merely serving as a relativistic extension of the former\cite{Peskin2018,Kleinert2009}. Given the broad applicability of path integrals in both classical quantum mechanics and QED, this separation suggests the potential existence of an undiscovered physical law hidden within the conventional structure of path integrals. This law, potentially crucial for spacetime theories, might hold the key to unlocking longstanding mysteries in quantum mechanics, such as the origins of spin, the nature of quantum entanglement, the phenomenon of single-electron interference, and the mechanisms behind superconductivity. Revealing this could profoundly transform our understanding of the quantum domain, offering a more unified and comprehensive theoretical framework that reconciles classical mechanics with quantum mechanics.

Histories, some researchers have identified this incompetence in path integral theories and tried to find out this hidden path integral formula. The Feynman checkerboard model provides an elegant approach to bridge the Dirac equation with particle stochastic trajectories\cite{feynman2010quantum}. It ingeniously derives the one-dimensional Dirac equation from the zigzag motion of particles at the speed of light\cite{HENNEAUX1982127,TJacobson_1984,PhysRevLett.89.250403}. However, extending this model to higher dimensions encounters several challenges, including computational complexity, divergences in calculations, and maintaining gauge invariance\cite{ord1992feynman,kull2019feynman}. Despite efforts by researchers to address these issues using alternative summation rules\cite{bialynicki1994weyl,meyer1996quantum,KULL2002147,Kull:1999aa}, the method of tranfer matrix\cite{ORD1993244} and the Monte Carlo method\cite{Earle1996}, a universal theory applicable to arbitrary dimensions has not yet emerged. B. Gaveau and Roberto Quezada attempted to correlate higher-dimensional Dirac equations with stochastic paths via momentum space\cite{Gaveau1993,Gaveau1994}, necessitating the further clarification in physical meanings and interpretations. To further advance research in this field, Gian Fabrizio and colleagues proposed a path integral representation specifically tailored for Dirac particles in electromagnetic fields\cite{Angelantonj1995,Angelantonj1996,Angelantonj1998,Fradkin1990}. However, the question of how to extend this method to more general external field scenarios while maintaining mathematical and physical consistency remains an open issue\cite{,Alexandrou2000,Corradini2020}. Moreover, these studies significantly diverge in philosophy from the classic Feynman path integrals and fail to establish a cohesive relations with them. These obstacles underscore the difficulty to construct the inherent link between the Dirac equation and the foundational principles established by Feynman.

The decline in research focus within this domain as we progressed into the 21st century can be attributed to a perceived completeness of existing theories and a shift towards the practical applications of quantum technology. However, the challenges faced in quantum mechanics applications over the past two decades necessitate a renewed focus on this research. The absence of a comprehensive description of the spacetime mechanism for quantum non-local correlations in current quantum mechanics limits our ability to manipulate quantum properties freely in the field of quantum information. Additionally, prevailing theories inadequately address phenomena such as superconductivity, underscoring the imperative for a novel theoretical construct. A theory of space-time correlation, harmonizing classical Feynman path integrals with QED, holds the potential to illuminate these quantum phenomena and enhance our capacity to interpret and manipulate the quantum world, catalyzing technological innovations and paving the way for groundbreaking applications in quantum computing, superconductor, and beyond.  

\section{The Missing Path Integral and Its Potential Form}
The existence of a missing content in current path integrals is evidenced by their formal logical inconsistencies and incompleteness, which fail to encompass all aspects of quantum mechanics. We elucidate these points in the following text.
\subsection{Inconsistency in Classical Path Integral Form}
The classical path integral theory, or non-relativistic Feynman path integral theory, describes the evolution of single particles. Within this theoretical framework, the classical action associated with various particle trajectories is introduced into quantum mechanics as a phase factor. The propagator in quantum mechanics can be represented as a superposition of all these phase factors:
\begin{equation}
	K(\vec{r},t;\vec{r}_0,t_0 )=C_0^{\frac{n}{2}}\sum_{\wp_k}\exp(\mathrm{i}S_c/\hbar),
	\label{prop Eq}
\end{equation} 
where $S_c$ is the classical action and $C_0=m_0/2\mathrm{i}\pi(t-t_0)$ is a path-independent constant\cite{Feynman1948}. $\wp_k$ here is used to denote the function of the $k$-th path. However, this formulation presents challenges. The action $S_c$ in Eq.~\ref{prop Eq} is classical, yet the paths considered are arbitrary, encompassing those approaching or surpassing light speed. This raises a critical question regarding the consistency of the theory: is it feasible to exclude these non-classical paths? The answer is negative. Eliminating these non-classical paths would lead to the inability to derive the Schr\"odinger equation and would detach the path integral from its fundamental connection to quantum mechanics\cite{Wen_2012,Schulman1981}. 

Interestingly, contrary to intuitive understanding, within a time-sliced evolution process, which is a method to analyze quantum processes step-by-step in time, the contribution of classical path actions is marginal, with non-classical paths assuming a dominant role. This leads to another critical question: why rely on a classical action when these non-classical paths are so crucial? One might consider using a relativistic action in Eq.~\ref{prop Eq} as an alternative. However, substituting $S_c$ with a relativistic action would cause the formula to diverge, thereby rendering the path integral ineffective\cite{Wen_2012}.

This dilemma underscores a fundamental inconsistency within the classical path integral paradigm: the path selection is rooted in non-classical trajectories, yet the action is constrained within a classical mechanical quantity. This inconsistency results in the path integral lacking self-consistency and also implies that the classical path integral theory may simply be a projection of a more fundamental theory at low energies. It is an approximate theory. 

\subsection{Incompleteness of Path Integral Forms}
In the current quantum theory, it is distinguished by two principal path integral formulations. The first pertains to the classical action's relationship with the Schr\"odinger equation, providing a foundational framework for non-relativistic quantum mechanics. Conversely, the second formulation delves into the complex dynamics between Feynman diagrams and the Dyson series, essential for understanding quantum field theory and QED. Despite these advancements, a significant gap has emerged: there is no path integral formulation that relates to the Dirac equation.

The Dirac equation stands as a cornerstone of relativistic quantum mechanics, offering profound insights into the quantum nature of spin with a value of $1/2$ and elucidating complex phenomena such as the fine structure of hydrogen atom energy levels and the Land\'e $g$-factor. The expectation for a comprehensive path integral theory to illuminate these quantum mechanical nuances is both necessary and pressing. Yet, the current theoretical framework lacks a path integral model that seamlessly integrates the Dirac equation with relativistic action, thus limiting our understanding of these phenomena from a spacetime perspective.

This gap in the theoretical foundation not only hinders our ability to fully comprehend the underlying principles driving these phenomena but also suggests that the existing path integral theory is incomplete. Without a model that aligns with the Dirac equation, path integral theory remains a supplemental, albeit crucial, aspect of quantum mechanics rather than embodying a fundamental and comprehensive set of physical principles. This oversight underscores the urgent need for theoretical expansion to bridge this divide, promising to elevate our understanding of quantum mechanics to a more profound and unified level.

\subsection{The Missing Path integral and Its Potential Form}
From the above analysis, it is apparent that in quantum mechanics, there exists a missing path integral. In previous work, some researchers tried to find out its form. However, these forms either still belong to the path integrals under field theory, are restricted path integrals, or are path integrals with modified action, and have not revealed the true missing path integral form. We posit that the sought-after form of this path integral must adhere to the following three critical criteria: 1) \textbf{Alignment with the Dirac Equation}: This requirement is foundational, as the Dirac equation is pivotal in relativistic quantum mechanics; 2) \textbf{Compatibility with Existing Path Integral Formulations}: This criterion ensures the new form's coherence and integration within the broader theoretical framework of quantum mechanics; 3) \textbf{Convergence and Integrability}: The property that path integral forms must be convergent and integrable is the mathematical requirement for a physical theory and a prerequisite for ensuring its theoretical soundness and applicability.

It is important to note that to make the path integral form a non-diagonalized spinor form (a diagonalized spinor form is a trivial form), it is necessary to strip the path integral of its propagator identity. For an $n$-dimensional wave function, $\Psi=(\psi_1,\psi_2,\cdot,\psi_n)^{T}$, it satisfies
\begin{align}
 	\Psi(\vec{r},t)&=\left(
 	\begin{matrix}
 		\langle \vec{r},t|\psi_1\rangle, &
 		\langle \vec{r},t|\psi_2\rangle, &
 		\cdots, &
 		\langle \vec{r},t|\psi_n\rangle
 	\end{matrix}
 	\right)^T \nonumber \\
 	&=\int 
 	\mathrm{diag}\left(
 	\begin{matrix}
 		\langle \vec{r},t|\vec{r}_0,t_0\rangle \\
 		\langle \vec{r},t|\vec{r}_0,t_0\rangle\\
 		\vdots \\
 		\langle \vec{r},t|\vec{r}_0,t_0\rangle
 	\end{matrix}
 	\right)^{T}
 	\left(
 	\begin{matrix}
 		\langle \vec{r}_0,t_0|\psi_1\rangle \\
 		\langle \vec{r}_0,t_0|\psi_2\rangle \\
 		\vdots \\
 		\langle \vec{r}_0,t_0|\psi_3\rangle
 	\end{matrix}
 	\right) \mathrm{d}^{3}\vec{r}_0. \nonumber\\
 &=\int K_{diag}(\vec{r},t;\vec{r}_0,t_0)\Psi(\vec{r}_0,t_0)\mathrm{d}^3\vec{r}_0\nonumber \\
 &=\int K_{diag}\hat{M}(\vec{r}_0,t_0)\Phi(\vec{r}_0,t_0)\mathrm{d}^3\vec{r}_0 \nonumber\\
 \Rightarrow \Phi(\vec{r},t)&=\int \hat{M}^{-1}(\vec{r},t)K_{diag}\hat{M}(\vec{r}_0,t_0)\Phi(\vec{r}_0,t_0)\mathrm{d}^3\vec{r}_0 \nonumber\\
 &=\int \hat{K}_{non-diag}\Phi(\vec{r}_0,t_0)\mathrm{d}^3\vec{r}_0
 \label{operator-K}
\end{align}
where $\hat{M}$ is the operator corresponding to the unitary transformation matrix and $\mathrm{diag}(a_1,a_2,\cdots,a_n)$ denotes the diagonal matrix with diagonal elements $a_1,a_2,\cdots,a_n$. Operator $\hat{M}$ transforms the time-evolved wave function $\Psi(\vec{r}_0,t_0)$ under diagonalization into $\Phi(\vec{r},t_0)$. The difference between $\Psi(\vec{r},t_0)$ and $\Phi(\vec{r},t_0)$ is that the temporal evolution of the component functions of $\Psi(\vec{r},t_0)$ is independent, i.e., $\psi_m(\vec{r},t)=e^{\mathrm{i}\int_{t_0}^{t}\hat{H}\mathrm{d}t_1/\hbar}\psi_m(\vec{r},t_0)$, whereas the temporal evolution of the component functions of $\Phi(\vec{r},t_0)$ is not independent, i.e., $\phi_m(\vec{r},t)=\sum_{k=0}^{n}a_k e^{\mathrm{i}\int_{t_0}^{t}\hat{H}\mathrm{d}t_1/\hbar}\phi_k(\vec{r},t)$. Eq.~\ref{operator-K} tells us that the path integral theory of non-diagonalized spinor form must necessarily be an operator. To establish the path integral theory linking relativistic action with the Dirac equation in spinor form, we must abandon the notion that the formulation of path integral theory must be a pure function, strip it of its propagator identity, and build its general form from the functional operator perspective.

Since this path integral form needs to include the relativistic action and must be an operator, we can express this path integral in a general form:
\begin{equation}
K_R = \hat{K}_{non-diag}=\hat{R} \sum_{\wp_k} e^{\mathrm{i}S_{R}(\wp_k)/\hbar},
\end{equation}
where $K_R$ represents the spinor form of the relativistic path integral, $S_R$ is the relativistic action in spinor form, which is different from the scalar relativistic action $S_r=\int_{t_0}^{t}(-\gamma m_0c^2-q\vec{A}+qV)\mathrm{d}t_1$. $\gamma$ represents the Lorentz factor, equaling $1/\sqrt{1-v^2/c^2}$. $\hat{R}$ is an operator that is independent of the path.

There might be reservations about $\hat{R}$ being in an operator form. However, transitioning $\hat{R}$ to an operator form is not merely an inevitable result of evolving from diagonalized propagators to non-diagonalized spinor path integrals; it also signifies our mathematical requirement to move from handling the evolution of plane waves to accommodating general spinor wavefunctions (as we will see in subsequent sections). This insight further stimulates our contemplation on the path integral form. In the context of Feynman's path integral theory, the normalization factor $C_0$ in Eq.~\ref{prop Eq} is traditionally considered path-independent, a premise that, while widely accepted, lacks rigorous theoretical or empirical validation. This raises an intriguing possibility that the constancy of $C_0$ might be an effective approximation within a more generalized path integral framework. Such a perspective points to a new clue for us to explore a path integral that transcends the limitations of current formulations, particularly by ensuring that $\hat{R}$ smoothly converges to $C_0$ in the low-energy regime, thereby maintaining compatibility with the Feynman path integral. Importantly, despite the spinor form of the path integral described in Eq.~\ref{operator-K} losing its identity as a propagator, it still functions as a spacetime entity that delineates the evolution of the wave function, thereby establishing a link between the action and the quantum mechanics evolution equation. What are the expressions for $\hat{R}$ and the spinor action $S_R$, and what physical insights do these expressions embody? These questions will be progressively addressed in the following discussions.
   
\section{Construction of Path Integral Theory in Spinor Form}
\subsection{Tricomi function and the spinor path integral}
The key to constructing the path integral in spinor form is to determine the expressions for $\hat{R}$ and $S_R$. The form of $S_R$ is crucial; choosing an appropriate expression for $S_R$ can simplify the expression of $\hat{R}$ and make its physical meaning more apparent.

The core issue in determining $S_R$ is how to write $\gamma m_0c^2$ in spinor form. Historically, some researchers have conducted studies on this, but influenced by Dirac's reformulation of the Klein-Gordon equation, it is generally believed that  $\sqrt{1-v^2/c^2} = \beta \pm \vec{\alpha} \cdot \vec{v}/c$. Here $\vec{\alpha}$ and $\beta$ use the Dirac representation\cite{Sakurai2017}. However, this spinor expression is not appropriate. Due to the specificity of spinor forms, $\beta m_0c^2-m_0c\vec{\alpha}\cdot\vec{v}$ has different diagonalization matrices for different $\vec{v}$ values, making it difficult to construct the connection $\beta m_0c^2\pm m_0c\vec{\alpha}\cdot\vec{v}\rightarrow \beta m_0c^2+\vec{\alpha}\cdot\hat{\vec{p}}c$ in the path integral process.

To establish the connection between the relativistic action and the Dirac equation, it is required that $\sqrt{1-v^2/c^2} = \beta  - \beta \vec{\alpha} \cdot \vec{v}/c$. We must note that this spinor form belongs to $Cl_{3,1}$, making $|\vec{v}|<c$ and $|\vec{v}|>c$ belong to different branches. To ensure the convergence of the integral for $|\vec{v}|>c$, we require $\sqrt{1-v^2/c^2}=\mathrm{i}^{\mathrm{sgn}(t_0-t)}\sqrt{v^2/c^2-1}$. Consequently,
\begin{equation}
	S_R(\wp)=\int_{\wp}(-\beta m_0c^2+\beta m_0c\vec{\alpha}\cdot\vec{v}-qV+q\vec{A}\cdot\vec{v})\mathrm{d}t_1.
\label{S_R express}
\end{equation}
$V$ and $\vec{A}$ here are used to denote the scalar and vector potential respectively. In this new expression, the diagonalization matrix $L_0$ of $S_R$ can be obtained:
\begin{align}
	L_0 &= e^{-\frac{1}{2} \mathrm{arctanh}(\vec{\alpha}\cdot \vec{v}/c)}=\sqrt{\frac{\gamma+1}{2}}- \sqrt{\frac{\gamma-1}{2}}\vec{\alpha}\cdot\vec{n}_v\nonumber \\
		&=\left(
			\begin{matrix} 
				\sqrt{\frac{\gamma+1}{2}}I & -\sqrt{\frac{\gamma-1}{2}}\vec{\sigma}\cdot\vec{n}_v \\
				-\sqrt{\frac{\gamma-1}{2}}\vec{\sigma}\cdot\vec{n}_v & \sqrt{\frac{\gamma+1}{2}}I 
			\end{matrix} 
		\right),
\label{L0 express}
\end{align}
where $\vec{n}_v=\vec{v}/|\vec{v}|$ is the unit vector of $\vec{v}$, and $\vec{\sigma}$ signifies the Pauli matrices. Although the expression of $L_0$ is not often seen in physics, it is in fact a Lorentz transformation under the structure of the Clifford algebra. Under the Clifford algebra structure, spacetime can be expressed as $\mathbbm{R}=ct+\vec{\alpha}\cdot\vec{r}$, and under this expression, the spacetime transformation (Lorentz transformation) can be written as:$\mathbbm{R}'=L_0^{-1}\mathbbm{R}L_0$. 

According to Eq.~\ref{S_R express}, we can see the eigenvalues of $S_R$ contain terms $\pm m_0c^2\sqrt{1-v^2/c^2}$. Thus, using the path integral in spinor form to calculate the evolution of the wave function will inevitably involve the following integral form:
\begin{equation}
\int_{-\infty}^{\infty} f(v) e^{\pm \mathrm{i}\chi\sqrt{1-\frac{v^2}{c^2}}} \phi(x_0 + v\varepsilon, t) \mathrm{d}(v\varepsilon )
\end{equation}
where $\chi=m_0c^2\varepsilon/\hbar$ is dimensionless constant and $f(v)$ is an integral function brought about by $\hat{R}$ and the diagonalization matrix $L_0$. This integral is quite complex and generally does not have an analytical solution. However, we notice that
\begin{equation}
	\begin{aligned}
		&\int_{-\infty}^{\infty} \gamma^{1/2}\sqrt{\frac{\gamma\pm 1}{2}} e^{\pm\mathrm{i}\chi\sqrt{1-\frac{v^2}{c^2}}} e^{\mathrm{i}pv\varepsilon} \mathrm{d}v  \\
		\sim &\sum_{k}a_k(p)U(-k,\frac{1}{2}-2k,\mp2\mathrm{i}\chi),
	\end{aligned}
\label{integral-Tricomi}
\end{equation}
The function $ U(a,b,c) $ is known as the Tricomi function. When $ a $ is a non-positive integer, it is closely related to the spherical Bessel functions $ j_a(1/c) $ and $ y_a(1/c) $\cite{Olver2010}. It is known that the spherical Hankel functions $ h_k^{\pm} = j_k(x) + \mathrm{i}y_k(x) $\cite{AbramowitzStegun1972} have the generating function $ \frac{1}{z} e^{\pm\mathrm{i}\sqrt{z^2-2zt}} $\cite{Arfken2013}. Hence, we get the following relationship:
\begin{equation}
	\begin{aligned}
	&\sum_{k}a_k(p)U(-k, \frac{1}{2} - 2k, \pm 2\mathrm{i}\chi) \\
	\sim &\sum_{k}b_k(p)h_k^{\pm}(\chi) \sim e^{\pm\mathrm{i}\chi\sqrt{1+\left(\frac{p}{p_0}\right)^2}}.
	\end{aligned}
\label{Tricomi-Dirac}
\end{equation}
This relationship is crucial for the construction of relativistic path integrals, implying that the integral with a core of $e^{\mathrm{i}L_r\varepsilon/\hbar}$ can be transmuted into the function $e^{-\mathrm{i}\hat{H}_r\varepsilon/\hbar}$, thus making it possible for us to establish the linkage between relativistic classical mechanics and relativistic quantum mechanics. The essential element in forging this link, namely the expression for $\hat{R}$, can be ascertained through the relationship between the Tricomi functions and the generating functions of the Hankel functions. After calculation, we come to these very strong conclusions as follows
\begin{equation*}
	 \begin{aligned}
 	 & \int_{-\infty}^{\infty}\left(\frac{2\gamma}{1+\gamma}\right)^{\frac{n-1}{2}}\sqrt{\frac{\gamma^2 +\gamma}{2}}e^{\pm\mathrm{i}S_r/\hbar}\psi(\vec{r}_0,t_0)\mathrm{d}^n\vec{r}_0 \\
 	&=\mathrm{i}^{\frac{n}{2}\mp\frac{n}{2}}C_0^{-\frac{n}{2}}\left(\frac{2\hat{\gamma}}{1+\hat{\gamma}}\right)^{\frac{n-1}{2}}\sqrt{\frac{\hat{\gamma}^2 +\hat{\gamma}}{2}}e^{\mp\mathrm{i}\hat{H}_r\varepsilon/\hbar}\psi(\vec{r},t_0), \\
 	& \int_{-\infty}^{\infty}\left(\frac{2\gamma}{1+\gamma}\right)^{\frac{n-1}{2}}\sqrt{\frac{\gamma^2-\gamma}{2}}\frac{\vec{v}}{|\vec{v}|}e^{\pm\mathrm{i}S_r/\hbar}\psi(\vec{r}_0,t_0)\mathrm{d}^n\vec{r}_0 \\
 	&=\mathrm{i}^{\frac{n}{2}\mp \frac{n}{2}\pm1}C_0^{-\frac{n}{2}}\left(\frac{2\hat{\gamma}}{1+\hat{\gamma}}\right)^{\frac{n-1}{2}}\sqrt{\frac{\hat{\gamma}^2-\hat{\gamma}}{2}}\frac{\hat{\vec{v}}}{|\hat{\vec{v}}|}e^{\mp\mathrm{i}\hat{H}_r\varepsilon/\hbar}\psi(\vec{r},t_0), 
 \end{aligned}
\end{equation*}
where, $\hat{\vec{v}}=(\hat{\vec{p}}-q\vec{A})/(\beta m_0)$, and $\hat{\gamma}=1/\sqrt{1-\hat{\vec{v}}^2/c^2}$. $\hat{H}_r=\sqrt{m_0^2c^4+(\hat{\vec{p}}-q\vec{A})^2}+qV$ is a scalar Hamiltonian in relativity. $\mathrm{d}^n\vec{r} $ is the abbreviation of the form $\mathrm{d}x_1\mathrm{d}x_2\cdots\mathrm{d}x_n$. The details of these calculations are provided in supplementary material. Basing these conclusions above, we can construct the expression for $\hat{R}$ under $n$-D like follows
\begin{align}
	&\hat{R}(\vec{r},\vec{r}_0;t,t_0)=L_n^{-1}(\hat{\vec{v}})\left(\frac{m_0\beta}{2\mathrm{i}\pi\hbar(t-t_0)}\right)^{\frac{n}{2}}L_n(\vec{v}), \label{R_expression}\\
	&L_n(\vec{v})=\gamma^{\frac{1}{2}}\left(\frac{2\gamma}{1+\gamma}\right)^{\frac{n-1}{2}} L_0.
	\label{Ln-expression}
\end{align}
Here, $\vec{v}=(\vec{r}-\vec{r}_0)/(t-t_0)$ and $\gamma(\hat{\vec{v}})=\hat{\gamma}$. It should be emphasized that the expression for $\hat{R}$ is unique. As can be seen from Eq.~\ref{integral-Tricomi}, changing the expression of $\hat{R}$ will change the value of $a_k(p)$, which will result in the resulting expression for the path integral being linearly independent of $e^{\pm\mathrm{i}\chi\sqrt{1+(p/p_0)^2}}$, and thus it will be impossible to construct a relationship between $S_r$ and $\hat{H}_r$.

Incorporating the expressions for $S_R$ in Eq.\ref{S_R express} and $\hat{R}$ in Eq.\ref{R_expression}, we have
\begin{align}
	&\begin{cases}
		K_R\left(\vec{r}, t ; \vec{r}_0, t_0\right)=\hat{R}\sum_{\wp_k} \exp\left(\frac{\mathrm{i}S_R(\wp_k)}{\hbar}\right)  \\
		\Psi(\vec{r},t)=\int_{-\infty}^{\infty}K_R(\vec{r},t;\vec{r}_0,t_0)\Psi(\vec{r}_0,t_0)\mathrm{d}^{n}\vec{r}_0 
	\end{cases} 
	\label{KRexpression}
	\\
	\Rightarrow 
	&\begin{cases}
		\hat{H}_{R}=\beta m_0c^2+\vec{\alpha}\cdot(\hat{\vec{p}}-q\vec{A})c+qV(\vec{r},t)\\
		\Psi(\vec{r},t)=\hat{T}e^{-\frac{\mathrm{i}}{\hbar}\int_{t_0}^{t}\hat{H}_R\mathrm{d}t_1}\Psi(\vec{r},t_0) \\
		\mathrm{i}\hbar\frac{\partial}{\partial t}\Psi(\vec{r},t)=\hat{H}_R \Psi(\vec{r},t).
	\end{cases}
\end{align}
It is the Dirac equation.

\subsection{Exposition on the Spinor Form of Path Integral Expression}
From the analysis above, we can see that the expression for $\hat{R}$ is the result of mathematical operations. It may seem complicated, but holds substantial physical significance. Specifically, in the expression for $L_n$, $L_0 = e^{-\frac{1}{2}\mathrm{arctanh}(\vec{v}/c)}$ represents the Lorentz transformation, which is a physical quantity that must appear in the transition from a diagonal propagator to a spinor form of the path integral. Moreover, the occurrence of the term $\gamma^{n/2}$ in Eq.~\ref{Ln-expression} is notably deliberate, mirroring the relativistic adjustment of the original coefficient $C_0$ to accommodate proper time $\Delta\tau = (t-t_0)/\gamma$, a Lorentz invariant. This adaptation emphasizes the relativistic covariance of the path integral formulation. As for the term $\sqrt{(1+\hat{\gamma})/2}$, though less common in physics, aligns with the normalization factor $\sqrt{(H_r + m_0c^2) / 2H_r}$ for Dirac plane waves, underscoring its relevance in relativistic corrections. 

It is important to note that since the conclusion
\begin{equation}
\int_{t_0}^{t}\beta m_0c^2(1+\vec{\alpha} \cdot \frac{\vec{v}}{c})\mathrm{d}t_1 = \beta m_0c^2(t-t_0 + \vec{\alpha} \cdot \frac{\vec{r} - \vec{r}_0}{c})
\end{equation}
is path-independent, $K_R$ in Eq.~\ref{KRexpression} can be further simplified as:
\begin{align}
	K_R(\vec{r}, \vec{r}_0; t, t_0) = &\hat{R}e^{\frac{\mathrm{i}}{\hbar}(\beta m_0c^2(t-t_0) + \beta m_0 c\vec{\alpha} \cdot (\vec{r}-\vec{r}_0))} \nonumber\\
	&\sum_{\wp_k}e^{\frac{\mathrm{i}}{\hbar}\int_{\wp_k}(q\vec{A} \cdot \vec{v} - qV)\mathrm{d}t_1}\nonumber \\
	=& K_{0}(\vec{r}, \vec{r}_0; t, t_0)\sum_{\wp_k}e^{\frac{\mathrm{i}}{\hbar}\int_{\wp_k}(q\vec{A} \cdot \vec{v} - qV)\mathrm{d}t_1}.  \nonumber
\end{align}
Here, we use $K_0$ to represent the path integral in spinor form without potential energy. The path independence of the integration involving $\beta m_0c^2(1+\vec{\alpha} \cdot \frac{\vec{v}}{c})$ across time leads to a significant simplification in the expression for $K_R$. This simplification reveals that the path integral's dependence on particle paths, in the absence of potential energy boils down to the contributions from vector and scalar potentials. 

The integration of the vector potential $\vec{A}$ across different paths affects the kernel $K$, but intriguingly, it does not alter the wave function's evolution. Under different paths, the contribution of $\vec{A}$ to $K$ is expressed as: $e^{\frac{\mathrm{i}}{\hbar}\int_{\wp_k}q\vec{A}\cdot\vec{v}\mathrm{d}t_1} = e^{\frac{\mathrm{i}}{\hbar}(F_k(\vec{r}) - F_k(\vec{r}_0))}$. Here, $F_k$ satisfies $\nabla F_k(\vec{r}) = q\vec{A}$, which leads to $e^{\mathrm{i}F_k(\vec{r})/\hbar}\hat{\vec{p}}e^{-\mathrm{i}F_k(\vec{r})/\hbar } = \hat{\vec{p}}-q\vec{A}$. When using Eq.~\ref{KRexpression} to calculate the evolution of the state function, $e^{-\mathrm{i}F_k(\vec{r}_0)/\hbar}$ will be included in the integration. After integration, $e^{-\mathrm{i}F_k(\vec{r}_0)/\hbar}$ will become $e^{-\mathrm{i}F_k(\vec{r})/\hbar}$, forming the structure $e^{\mathrm{i}F_k(\vec{r})/\hbar}\hat{U}e^{-\mathrm{i}F_k(\vec{r})/\hbar } $ with the evolution operator $\hat{U} = \hat{T}e^{\frac{\mathrm{i}}{\hbar}\int_{t_0}^{t}\hat{H}_R\mathrm{d}t_1}$, eventually leading to $\hat{H}_R(\hat{p}) = \hat{H}_R(\hat{p} - q\vec{A})$. This analysis culminates in the realization that, despite the vector potential's variable contribution under different paths to $K_R$, its effect on the state function's evolution remains uniform. This uniformity, encapsulated in the transformation $e^{\mathrm{i}F_k(\vec{r})/\hbar}\hat{\vec{p}}e^{-\mathrm{i}F_k(\vec{r})/\hbar } = \hat{\vec{p}}-q\vec{A}$, ensures that the vector potential's integral translates identically across varying paths in the evolution of the state function.

In contrast, the scalar potential $V(\vec{r}, t)$ exhibits a true path dependency within the spinor form of the path integral. This dependency complicates the application of path integrals for analyzing the wave function's long-term evolution, confining our analysis to within a temporal slice $\varepsilon$. Within such a time slice, as $\varepsilon \rightarrow 0$, the sum $\sum_{\wp_k} e^{\frac{\mathrm{i}}{\hbar}\int_{\wp_k}V\mathrm{d}t_1}$ becomes proportional to $e^{\frac{\mathrm{i}}{\hbar}V(\vec{r},t)\varepsilon}$. Furthermore, as $\varepsilon \rightarrow 0$, the operator $\hat{\vec{p}} e^{\frac{\mathrm{i}}{\hbar}V(\vec{r},t)\varepsilon}$ approaches $e^{\frac{\mathrm{i}}{\hbar}V(\vec{r},t)\varepsilon}\hat{\vec{p}}$. Since $e^{\frac{\mathrm{i}}{\hbar}\int_{\wp_k}V\mathrm{d}t_1} = e^{\frac{\mathrm{i}}{\hbar}(f_k(t) - f_k(t_0))}$, leading to
\begin{align}
	&\Psi(\vec{r},t)=\int_{-\infty}^{\infty}K_0e^{\frac{\mathrm{i}}{\hbar}(f_k(t_0)-f_k(t))}\Psi(\vec{r}_0,t_0)\mathrm{d}^{n}\vec{r}_0 \nonumber \\
	\Rightarrow & e^{\varepsilon\partial_{t_0}}\Psi(\vec{r},t_0)= e^{\frac{\mathrm{i}}{\hbar}f_k(t)}e^{-\varepsilon\frac{\mathrm{i}}{\hbar}(\beta m_0c^2+\vec{\alpha}\cdot\hat{\vec{p}}c)}e^{\frac{\mathrm{i}}{\hbar}f_k(t_0)}\Psi(\vec{r},t_0) \nonumber \\
	\Rightarrow & e^{-\frac{\varepsilon\mathrm{i}}{\hbar}(\mathrm{i}\hbar\partial_{t_0}-V)}\left(e^{\frac{\mathrm{i}}{\hbar}f_k}\Psi\right)=e^{-\frac{\varepsilon\mathrm{i}}{\hbar}(\beta m_0c^2+\vec{\alpha}\cdot\hat{\vec{p}}c)}\left(e^{\frac{\mathrm{i}}{\hbar}f_k}\Psi\right). \nonumber
\end{align}
This relationship holds for any wave function $\Psi(\vec{r},t_0)$, and thus 
\begin{equation}
	(\mathrm{i}\hbar\partial_{t}-V)\Psi'(\vec{r},t)=(\beta m_0c^2+\vec{\alpha}\cdot\hat{\vec{p}}c)\Psi'(\vec{r},t).
\end{equation}
where $\Psi'(\vec{r},t)=e^{\frac{\mathrm{i}}{\hbar}f_k(t)}\Psi(\vec{r},t)$. It should be emphasized that the derivation of the Dirac equation involving the scalar potential is based on the premise that $\varepsilon \rightarrow 0$. Without this premise, $\hat{\vec{p}}$ would not have a commutation relation with $e^{\frac{\mathrm{i}}{\hbar}\int_{\wp_k}V\mathrm{d}t_1}$, causing the operator $\hat{p}_i$ in the Hamiltonian to become $\hat{p}_i + V/v_i$, which is inconsistent with the Dirac equation. It is precisely because of the presence of the scalar potential $V$ that the dynamical evolution equation (Dirac equation) replaces $\Psi(\vec{r},t)=e^{\frac{\mathrm{i}}{\hbar}\int_{t_0}^{t}\hat{H}_R\mathrm{d}t_1}\Psi(\vec{r},t_0)$ as the most fundamental principle in quantum mechanics.

\section{Relationship between Spinor Form of Path Integrals and other forms of path integrals}
\subsection{Spinor Form of Path Integrals versus Scalar Form of Path Integrals}

In previous work, Wen, etc., established the path integral in scalar form, constructing the relationship between the relativistic Lagrangian in scalar form, $L_r = \gamma m_0c^2 + q\vec{A} - qV$, and the scalar form of the Hamiltonian, $\hat{H}_r = \sqrt{m_0^2c^4 + (\hat{\vec{p}} - q\vec{A})^2c^2} + qV$. In fact, the path integral in spinor form is a further extension of this work. The diagonal form of the path integral in spinor form is essentially the path integral theory in scalar form. The main physical quantities causing the non-diagonalization of the path integral in spinor form are $L_0$ and $e^{\mathrm{i}S_R/\hbar}$. If we take the diagonal elements of these quantities, then we obtain
\begin{equation*}
\mathrm{diag}(\hat{R}\sum_{\wp_k}e^{\mathrm{i}S_R/\hbar})=
\left(
\begin{matrix}
	R \sum_i W e^{\frac{\mathrm{i} S_r}{\hbar}}, & 0\\
	0,& R \sum_i W e^{\frac{-\mathrm{i} S_r}{\hbar}}
\end{matrix}
\right)  
\end{equation*}
This is the expression for the path integral in scalar form that includes solutions with negative energy. This indicates that the path integral in scalar form is the result of decoupling the positive and negative energy state spaces in the spinor form of path integrals.

If we revisit the theory of path integrals in spinor form, we find that $S_R$ has lost its role in describing the motion of objects; we cannot derive the general laws of motion from the principle of least action. However, in the scalar form of path integral theory, $S_r$ still retains its physical significance as an action, capable of describing the classical motion of spin particles. The transformation from $S_r$ to $S_R$ requires us to introduce $-L_r$. In classical mechanics, we retained $L_r$ and discarded $-L_r$, but in the theory of path integrals, we need to include $-L_r$ into the new theory. 

From $L_R = L_0^{-1}(\beta L_r) L_0$, we know that because $\delta L_R$ involves the variation of $L_0$, $\delta\int L_R\mathrm{d}t = 0$ no longer describes the true trajectory of an object. This conclusion underscores that the scalar form of path integrals aligns more closely with Feynman's conceptualization of path integrals, embodying a more direct approach to quantum dynamics. Conversely, the spinor form, with $S_R$ detached from motion description, evolves into a theory of spacetime correlation. This theory, while not directly elucidating motion, offers a richer tapestry of insights, suggesting that the intertwining of positive and negative energy states underpins the fundamental laws of matter.

\subsection{Spinor Form of Path Integrals versus Classical Feynman Path Integrals}
From our analysis, we know that if the positive and negative energy state spaces are decoupled, then the spinor form of path integrals will transition to the scalar form. On this basis, if the quantum system's momentum $\langle (\hat{p} - q\vec{A})^2\rangle \ll m_0^2c^2$, implying $\langle\beta\hat{\vec{v}}\rangle \ll c$, this leads to $L_n(\hat{v}) \approx 1$, thus causing the operator $\hat{R}$ to degenerate into a path-independent constant. At this point, we obtain
\begin{equation*}
\mathrm{diag}(\hat{R}\sum_{\wp_k}e^{\mathrm{i}S_R/\hbar})\sim \left(
\begin{matrix}
	C_0^{\frac{n}{2}} \sum_i e^{\frac{\mathrm{i} S_c}{\hbar}}, & 0\\
	0,& C_0^{\frac{n}{2}} \sum_i e^{-\frac{\mathrm{i} S_c}{\hbar}}
\end{matrix}
\right) 
\end{equation*}
This is the classical Feynman path integral form that includes solutions with negative energy. Here $S_c=\int_{t_0}^{t}(\vec{p}^2/2m_0+q\vec{A}-V)\mathrm{d}t_1$. This tells us that the Feynman path integral is a low-energy approximation of the spinor form of path integrals when the positive and negative energy state spaces are decoupled.

\subsection{Spinor Form of Path Integrals versus Path Integrals in QED}

The path integral in spinor form and the Feynman path integral under QED belong to different levels of path integral theory. The former is a spacetime theory about the evolution of single particles, while the latter is based on the theory of particle creation and annihilation under field quantization. They are connected through the Dirac equation.

Before these two path integrals can be linked, the issue of covariance in the spinor form of path integrals needs to be addressed. It is noted that the Dirac equation derived from equation 11 is not covariant. In fact, this equation is also the initial equation that Dirac constructed based on the principle of correspondence from the relativistic Hamiltonian. Multiplying both sides of this equation by $\beta$ yields the standard form of the Dirac equation:
\begin{align}
	&(\beta(\mathrm{i}\hbar\partial_t-qV) - \beta\vec{\alpha}\cdot(\hat{\vec{p}}-q\vec{A})c-m_0c^2)\Psi=0 \nonumber \\
	\Leftrightarrow& (\gamma^{\mu}(\mathrm{i}\hbar\partial_u-qA_\mu)-m_0c)\Psi=0.
\end{align}
Where $\gamma^{\mu}$ are Dirac Matrixes\cite{Dirac1927,Sakurai2017}, $\partial_0=\partial/(c\partial t)$ and $A_0=V/c$. Mathematically, this equation and Eq.~\ref{KRexpression} are essentially two equivalent equations. However, this equation cannot be directly obtained from path integral theory, primarily because, in the single-particle path integral theory, time and space do not have equivalent status. In the single-particle path integral theory, the action involves an integration over time but not over space, while the propagator calculation involves spatial integration but not temporal integration. Therefore, the single-particle path integral theory is necessarily non-covariant. Then, does a covariant form of path integral theory exist? It is believed not to exist. In classical mechanics, a ``path'' refers to the ``trajectory'' of mechanical motion, involving ``changes in space over time'', which implies time is a variable, and space is a physical quantity, thus their statuses are different. Therefore, as long as the classical concept of ``path'' is retained, a covariant form of path integral does not exist. In fact, the ``path'' in Feynman path integrals under QED has deviated from the classical concept of ``trajectory''; it refers to ``types of scattering''. Precisely for this reason, the Feynman path integrals in QED do not possess a specific expression form but are merely a method for calculating the Dyson series.

\section{The Spinor Form of Path Integral Theory Reveals New Physical Laws}

\subsection{Spinor Form of Path Integrals and the Origin of Spin}
The origin of spin has long posed a big mystery in the realm of physics. From a mathematical perspective, the concept of spin space as the $2n$-dimensional inequivalent irreducible representation of the $SU(2)$ group is well-established\cite{Tinkham2003}. This mathematical formalism provides a solid foundation for describing the properties of spin, such as its quantization and transformation under rotations. However, the fundamental reasons for the emergence of spin and the specific details of its physical genesis remain subjects of considerable intrigue research.

Traditionally, the introduction of the $\vec{\alpha}$ matrices in the Dirac equation is regarded as a crucial step towards understanding spin. The Dirac equation, which combines quantum mechanics and special relativity, successfully describes the behavior of spin-$1/2$ particles, such as electrons. Yet, when examined within a broader algebraic context, the $\vec{\alpha}$ matrices merely serve as the fundamental vectors in the $Cl_{3,0}$ space, aligning with the inherent algebraic structure of Maxwell's equations\cite{spacetimealgebra,Lounesto:1986aa}. This observation suggests that the $\vec{\alpha}$ matrices alone do not fully capture the physical origin of spin.

To illustrate this point, let us consider Maxwell's equations, which govern the behavior of electromagnetic fields. If we define the vectors
\begin{align}
&\vec{r}=x\alpha_x+y \alpha_y+ z\alpha_z, \nonumber \\
&\vec{J}=J_x\alpha_x+J_y \alpha_y+ J_z\alpha_z, \nonumber \\
&\vec{E}=E_x\alpha_x+E_y \alpha_y+ E_z\alpha_z, \nonumber \\
&\vec{B}= B_x\alpha_x+B_y \alpha_y+ B_z\alpha_z, \nonumber
\end{align}
then Maxwell's equations can be succinctly expressed as \cite{spacetimealgebra}
\begin{equation}
(\frac{\partial}{c\partial t}+\vec{\nabla})(\vec{E}+\mathbbm{i}c\vec{B})=\sqrt{\frac{\mu}{\varepsilon}}(\rho c+\vec{J}).
\end{equation}
where $\mathbbm{i}=\alpha_x\alpha_y\alpha_z$ is the image number in $Cl_{3,0}$ space. This formulation reveals that, in describing Maxwell's equations, we can entirely introduce the $\vec{\alpha}$ matrices to reformulate the laws of electromagnetism, and this algebraic operation structure is completely equivalent to the modern vector description. In other words, the introduction of $\vec{\alpha}$ matrices, in itself, does not elucidate the origin of spin; had this been the case, spin as a physical quantity would have been apparent within the classical framework of Maxwell's equations.

From the perspective of spinor form path integrals, the key to making spin manifest in quantum mechanics is the introduction of the $\beta$ matrix, which shifts the physical laws from those belonging to the $Cl_{3,0}$ algebraic space of Maxwell's equations to the $Cl_{3,1}$ algebraic space of Dirac's theory\cite{Lounesto:1986aa}. The $\beta$ matrix, along with the $\vec{\alpha}$ matrices, forms the complete set of Dirac matrices, which are essential for describing the behavior of spin-1/2 particles. Notably, in our spinor path integral theory, the $\beta$ matrix always appears in conjunction with the mass $m_0$. This observation suggests that mass plays a crucial role in the emergence of spin.

The introduction of the concept of negative mass and allowing negative mass to form a spinor space with positive mass lays the foundation for constructing the Dirac equation. This mathematical framework enables the description of particles with both positive and negative energy states, which is a key feature of relativistic quantum mechanics. From this, we can speculate that spin might arise from the coupling of electromagnetic laws with particles of positive and negative mass. In other words, the interplay between the algebraic structure of electromagnetism ($Cl_{3,0}$) and the extended algebraic structure that includes both positive and negative mass ($Cl_{3,1}$) could be the physical origin of spin.

This viewpoint offers a new perspective on understanding the physical origin of spin, but it still requires further theoretical and experimental research to validate and refine. Future research should delve deeper into the role of mass in the emergence of spin, as well as the possible connection between the algebraic structures of electromagnetism and relativistic quantum mechanics. Additionally, investigating the behavior of spin in various physical scenarios, such as in the presence of strong gravitational fields or in the context of particle interactions, could provide valuable insights into the fundamental nature of spin.

\subsection{Spinor Form of Path Integrals and Quantum Non-local Correlations}
Due to path integrals being spacetime correlation functions, which describe the correlation of various spacetime points during the evolution of the wave function, path integrals can be used to reveal the reasons behind the formation of quantum non-local correlations and the manipulation of such correlations. In fact, in the work by Wen, \textit{et al}., the randomness, irreversibility, instantaneousness, and basis vector preference of scalar particle wave function collapse were explained through the scalar form of path integral theory, turning the axioms of quantum measurement into a corollary under this scalar form of path integral theory\cite{Wen_2012,Wen2023}. The spinor form of path integral theory we established in this work is a spinor extension of the scalar form of path integrals, capable of revealing the mysteries of spin particle wave function collapse and characteristics such as spin particle entanglement. We will elaborate on this work in future publications.

\section{Conclusions}
In this work, we discussed the inconsistencies and incompleteness present in current path integral theory. Given that path integral theory is an important theory in quantum mechanics, its flaws suggest that a more fundamental path integral theory, which could establish the connection between the relativistic action and the Dirac equation, is hidden within quantum mechanics. By analyzing the possible forms of this path integral, we found this hidden theory. Mathematical analysis indicates that the form of the path integral theory we discovered is unique. Unlike previous theories, our approach does not necessitate specific path constraints or modifications to the action, representing a natural extension of Feynman's path integrals. We have elaborated on the connections between our theory and scalar path integrals, Feynman path integrals, and path integrals in QED, offering novel perspectives on quantum spin and quantum non-local correlations. This work significantly advances our understanding of quantum mechanics, opening new avenues for exploration in the quantum information and computation, superconductor and so on.

\textit{Acknowledgements}: This work is supported by Supported by National Natural Science Foundation of China (Grant No. 11904099), Natural Science Foundation of Hunan Province of China (Grant No. 2021JJ30210), and Excellent Youth Program of Hunan Provincial Department of Education (Grant No. 22B0609).

\bibliographystyle{apsrev4-1}
\bibliography{ref}

\end{document}